\documentclass[prl,reprint,twocolumn,superscriptaddress,noshowpacs,notitlepage,longbibliography]{revtex4-1}%
\usepackage{graphicx,bm,times}
\usepackage{amsmath}
\usepackage{amsfonts}
\usepackage{amssymb}
\usepackage{color}
\usepackage{hyperref}
\hypersetup{
	colorlinks = true,
	allcolors = {blue}
}

\begin{document}
	
	\title{On the origin of the anomalous peak in the resistivity of TiSe$_2$}
	
	\author{Matthew D. Watson}
\email{mdw5@st-andrews.ac.uk}
\affiliation {SUPA, School of Physics and Astronomy, University of St. Andrews, St. Andrews KY16 9SS, United Kingdom}

\author{Adam M. Beales}
\affiliation {SUPA, School of Physics and Astronomy, University of St. Andrews, St. Andrews KY16 9SS, United Kingdom}
	
\author{Philip D. C. King}
\email{philip.king@st-andrews.ac.uk}
\affiliation {SUPA, School of Physics and Astronomy, University of St. Andrews, St. Andrews KY16 9SS, United Kingdom}

\begin{abstract}
Resistivity measurements of TiSe$_2$ typically show only a weak change in gradient at the charge density wave transition at $T_{CDW}\approx$ 200~K, but more prominently feature a broad peak at a lower $T_{peak}\sim$ 165~K, which has remained poorly understood despite decades of research on the material. Here we present quantitative simulations of the resistivity using a simplified parametrization of the normal state band structure, based on recent photoemission data. Our simulations reproduce the overall profile of the resistivity of TiSe$_2$, including its prominent peak, without implementing the CDW at all. We find that the peak in resistivity corresponds to a crossover between a low temperature regime with electron-like carriers only, to a regime around room temperature where thermally activated and highly mobile hole-like carriers dominate the conductivity. Even when implementing substantial modifications to model the CDW below the transition temperature, we find that these thermal population effects still dominate the transport properties of TiSe$_2$. 

\end{abstract}
\date{\today}
\maketitle

Phase transitions such as charge density waves (CDWs) are often first characterized by the observation of anomalies in resistivity measurements. Such phase transitions can influence the resistivity via Fermi surface reconstructions, changes in scattering rates, and/or a loss of free carriers due to the formation of an energy gap. However in TiSe$_2$, which exhibits a much-studied CDW-like phase transition at $T_{CDW}\approx$~200~K \cite{DiSalvo1976,Craven1978}, transport measurements show a highly unusual and non-monotonic temperature-dependence. Samples which are close to stoichiometry typically show $n$-type metallic-like behavior at low temperatures, followed by a prominent broad peak at $T_{peak}\sim$165~K, distinct from $T_{CDW}$, beyond which the resistivity decreases with increasing temperature in a semiconductor-like fashion, with a positive Hall coefficient at room temperature \cite{DiSalvo1976}. Surprisingly little change occurs at $T_{CDW}$, with at most a modest change in slope observed at $\approx$200~K \cite{DiSalvo1976,Taguchi1981}, even though the CDW involves changes to the band structure on energy scales as large as 100~meV \cite{Rossnagel2011}. The resistivity is known to be highly sensitive to the sample stoichiometry, and the observation of the anomalous broad peak at $T_{peak}\sim$165~K in resistivity measurements has sometimes been taken as an indicator of sample quality \cite{DiSalvo1976,DiSalvo1978,Levy1980,Taguchi1981}. On the other hand, some studies have interpreted the peak feature as a signature of the role of excitons in the CDW ordering \cite{Monney2009PhysicaB,Monney2010PRB,Monney2010NJP,Koley2014,Hildebrand2016}. Given the resurgence of interest in TiSe$_2$ in recent years, it is worthwhile to revisit the long-standing problem of its unusual transport properties \cite{Wilson1978,DiSalvo1978,Rossnagel2002,CampbellPaglione2018_arxiv}, aided by the availability of recent characterizations of its 3D electronic structure by angle-resolved photoemission spectroscopy (ARPES) \cite{Watson2019PRL}.

In this paper, we show that the overall temperature-dependence of the resistivity of TiSe$_2$, including the anomalous peak, can be reasonably reproduced without accounting for the CDW at all. Our model is very simple, but it captures the essential ingredients necessary to qualitatively understand the transport properties of TiSe$_2$. The unusual resistivity profile is mainly due to the temperature-dependent thermal populations of the 2D hole and 3D electron bands, which vary significantly since the band gap is only $\sim3\times{}k_BT$ at room temperature. The apparently anomalous transport behavior of TiSe$_2$ can thus be intuitively understood as a consequence of the thermal activation of mobile hole-like carriers above a characteristic temperature of $\sim$150~K, which sets the scale of $T_{peak}$. $T_{peak}$ thus represents a crossover from electron-dominated to hole-dominated transport properties, which also manifests as a sign change in the Hall coefficient at a similar temperature, as found in experiments. We show that changes in the band structure due to the CDW may modulate the resistivity below $T_{CDW}$ and shift $T_{peak}$ a little, but the qualitative understanding of the anomalous peak remains unchanged. We finally discuss the appropriate criteria for extracting $T_{CDW}$ from resistivity experiments, with implications for the determining the boundaries of phase diagrams based on TiSe$_2$.  

\begin{figure*}
	\centering
	\includegraphics[width=\linewidth]{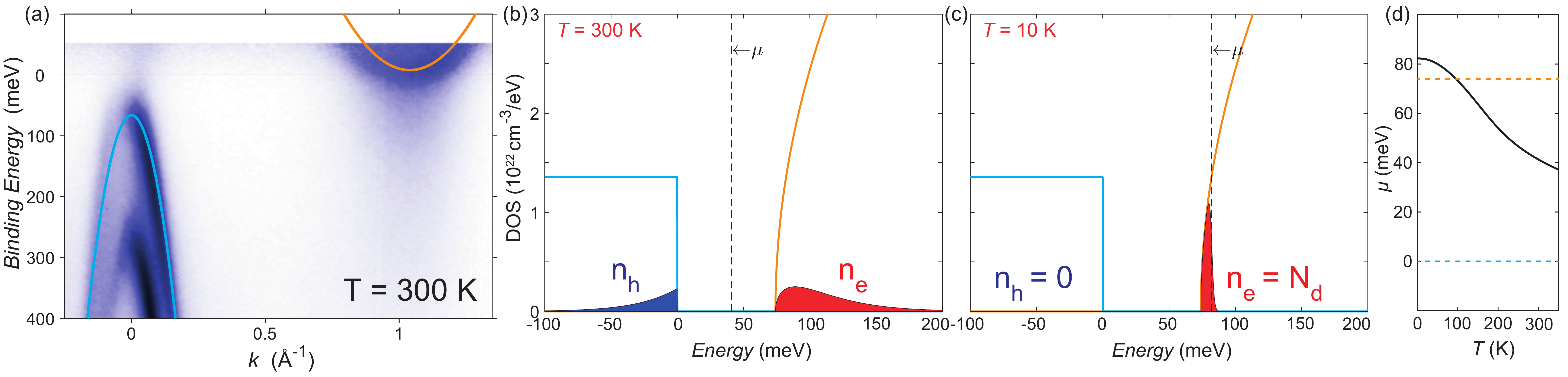}
	\caption{Model band structure of TiSe$_2$. (a) ARPES data at 300~K, adapted from \cite{Watson2019PRL}, overlaid with parabolas corresponding to the assumed dispersions of the hole and electron-like carriers respectively. (b) Available Density of States for the 2D hole band (cyan line) and 3D electron band (orange line), also showing the thermally occupied population of each carrier type (shaded areas) at 300~K, and (c) 10~K. The dashed line indicates the location of the chemical potential. (d) Chemical potential as a function of temperature. Note that in (b-d), the top of the valence band is defined to be at zero energy, whereas in (a) the data is referenced to the experimental chemical potential.}
	\label{fig:fig1}
\end{figure*}
\begin{figure}[t]
	\centering
	\includegraphics[width=0.9\linewidth]{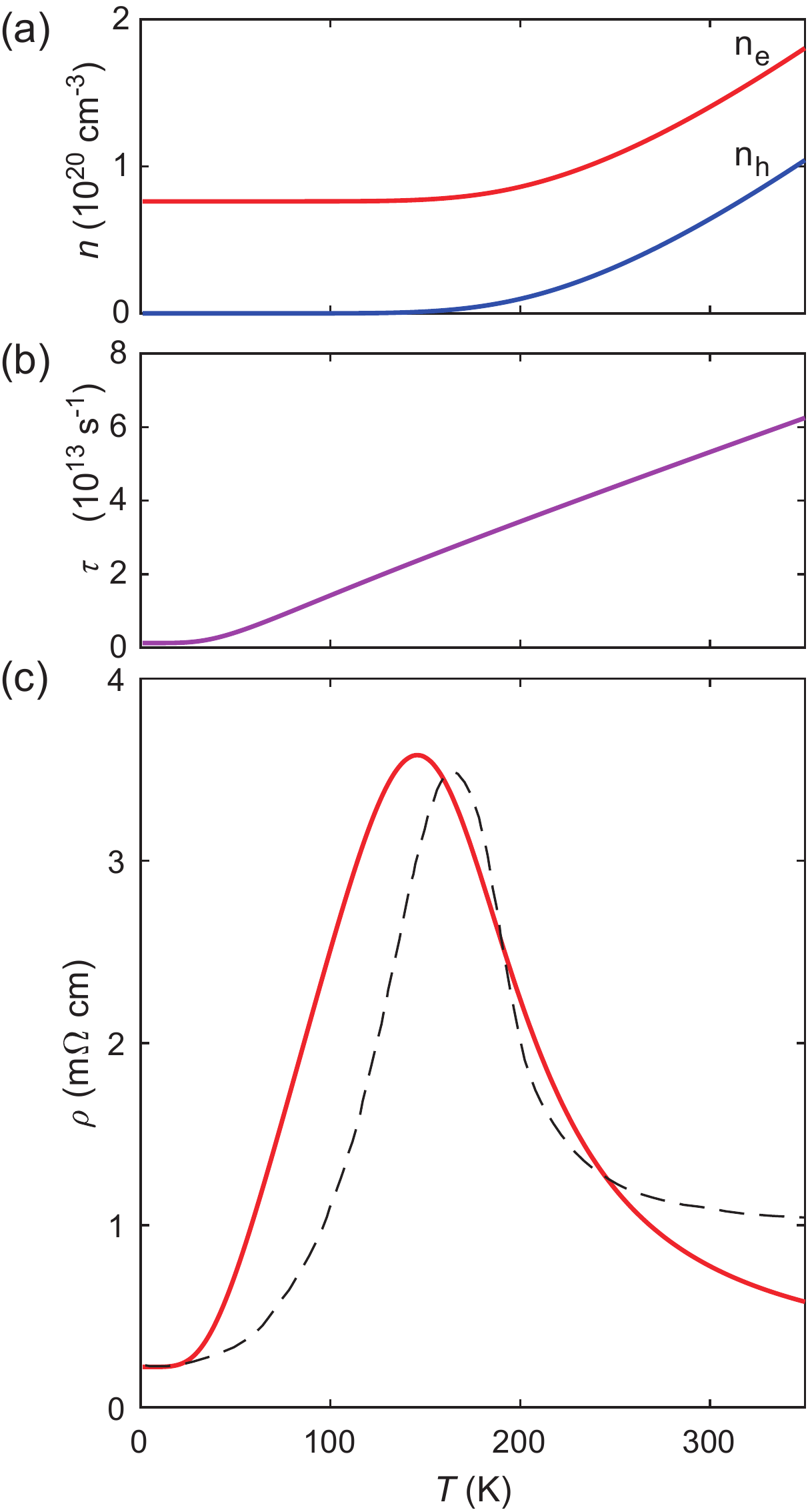}
	\caption{ (a) Temperature-dependence of electron- and hole-like carrier densities (see Fig.~\ref{fig:fig1}(b-c) for visualization at two selected temperatures). (b) The assumed scattering rates, following the Bloch-Gr\"{u}neisen formula for electron-phonon scattering with an additional elastic impurity scattering term. (c) Corresponding simulation of resistivity. Dashed line is the experimental resistivity, from Ref.~\cite{DiSalvo1976}.}
	\label{fig:fig2}
\end{figure}

\subsection{Model parameters and carrier densities}

Our model aims for a very simple description of transport in TiSe$_2$, but with all parameters constrained by experimental evidence. A crucial question from the outset is whether TiSe$_2$ should be treated as a semiconductor or a semimetal. Recent detailed ARPES measurements unambiguously observed a small band gap in the normal state \cite{Watson2019PRL}, and so we use the experimentally-determined band gap of 74 meV here. At the center of the Brillouin zone, TiSe$_2$ has two quasi-2D hole bands and one 3D hole-like band. We assume that the uppermost 2D hole band will dominate the contribution of hole carriers to transport, and we approximate this by a single 2D density of states (DOS), characterized by a single effective mass $m^*_{h}=$~0.31 $m_e$. The electron bands of TiSe$_2$ are centered at the L points on the edges of the Brillouin zone, and have a 3D anisotropic shape \cite{Watson2019PRL}. We approximate this with a 3D density of states, parametrized by a single scalar mass $m^*_{e}=$~3.8 $m_e$, and with a multiplicity factor of 3 due to the valley degeneracy. The effective mass parameters are set for agreement with the band dispersions observed by ARPES measurements in the normal phase \cite{Watson2019PRL}. Parabolas representing the assumed dispersions are plotted on top of ARPES data at 300~K in Fig.~\ref{fig:fig1}(a). TiSe$_2$ samples are never exactly stoichiometric, and to the best of our knowledge are always at least slightly $n$-doped, as indicated by the negative value of the Hall coefficient at low temperatures \cite{DiSalvo1976}. We therefore allow for a finite, temperature-independent, number of donors $N_{d}=7.62\times{}10^{19}$~cm$^{-3}$ into our model, which we fix such that $1/N_d|e|$ equals the low-temperature Hall coefficient reported in Ref.~\cite{DiSalvo1976}. 

While the model is clearly a simplification, we are careful to retain two key ingredients: the temperature-dependence of the chemical potential, and the corresponding thermal populations of hole and electron carriers \cite{Monney2009PhysicaB}. At each temperature, the chemical potential is determined by the requirement that $n_e(T)-n_h(T)=N_d$. This charge compensation condition leads to a downward shift of the chemical potential with increasing temperature (Fig.~\ref{fig:fig1}(d)). At low temperatures, in Fig.~\ref{fig:fig1}(c), there is essentially zero thermal population of the hole-like carriers, and $n_e=N_d$. However, by room temperature, since the band gap is only $\sim3\times{}k_BT$, the tails of the Fermi-Dirac state occupation function extend into the hole-like DOS and thermally populate a significant number of hole-like carriers, as shown in Fig.~\ref{fig:fig1}(b). This thermal population of hole-like carriers plays a leading role in the following simulations. 

\subsection{Simulation of resistivity}

We simulate $\rho(T)$, the resistance as a function of temperature, using the two-carrier Drude model:

\begin{equation}
\frac{1}{\rho(T)} = \frac{n_e(T)|e|^2\tau(T)}{m^*_e} + \frac{n_h(T)|e|^2\tau(T)}{m^*_h}
\end{equation}

The temperature-dependence of the electron and hole populations $n_{e,h}(T)$ is shown in Fig.~\ref{fig:fig2}(a). For the scattering time, $\tau{}(T)$, we take inspiration from the resistivity of TiTe$_2$, a sister material which is semimetallic, with a band overlap of several hundred millivolts. In Ref.~\cite{Allen1994PRB} it was shown that the resistivity of TiTe$_2$ followed the Bloch-Gr\"{u}neisen (B-G) formulation for scattering due to electron-phonon coupling. We therefore adopt this form of scattering rate for our model of TiSe$_2$: the coefficient of the B-G scattering term is a free parameter which we set for good agreement on the absolute value of resistivity around $\rho_{peak}$, and the Debye temperature of the B-G term is set to 250~K \cite{Craven1978,CampbellPaglione2018_arxiv}. We also add a constant scattering term from elastic impurity scattering, which is set such that $\rho^{exp}_{T\rightarrow{}0}=m_e^*/N_e|e|^2\tau_0$. 
Notably, this scattering rate, plotted in Fig.~\ref{fig:fig2}(b), is almost linear in temperature across an appreciable range, and featureless throughout the region of interest. An important assumption we make is that this scattering rate is identical for both electron and hole-like carriers. Importantly, however, the two types of carrier will have quite different mobilities \cite{Rossnagel2002}, since $\mu_{e,h}=|e|\tau/m^*_{e,h}$, and the effective masses of the electrons are much higher than the holes. 

\begin{figure*}
	\centering
	\includegraphics[width=\linewidth]{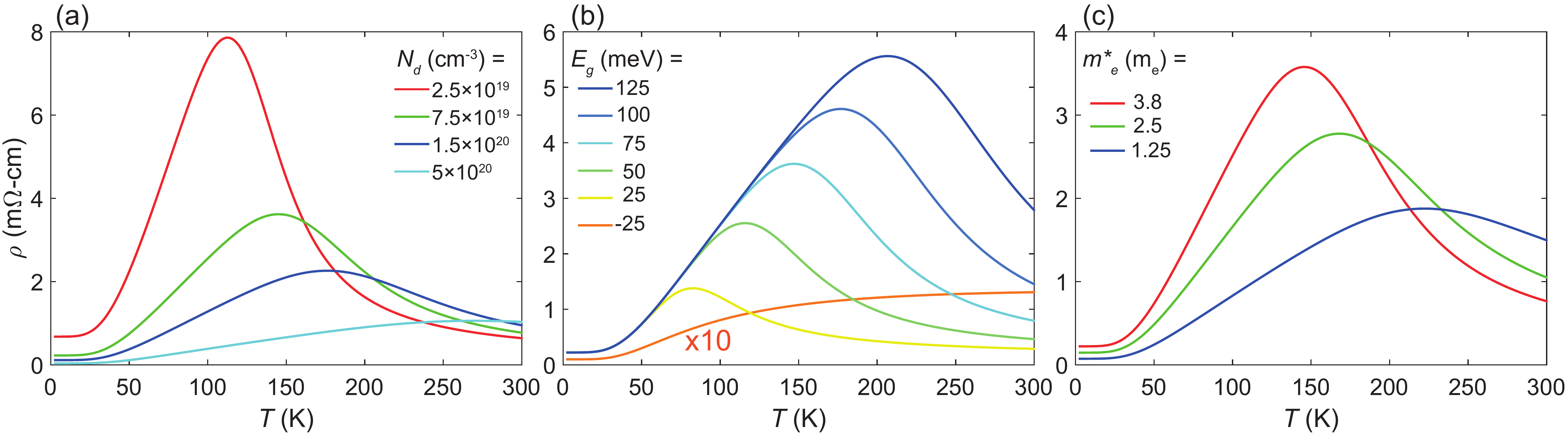}
	\caption{(a) Simulated resistivity curves for various values of the extrinsic doping $N_d$. All other parameters are held fixed, including the elastic impurity scattering term, which in reality would be likely to scale with $N_d$. (b) Variation of the resistivity with the band gap. Note that for the case of the band overlap, the resistivity is multiplied by a factor of 10 for greater visibility. (c) Simulations with various assumed values of the electron mass.}
	\label{fig:fig3}
\end{figure*}

 The simulated $\rho{}(T)$ is shown in Fig.~\ref{fig:fig2}(c), plotted in comparison to the experimental literature data from Di Salvo et al. \cite{DiSalvo1976} \footnote{We compare our resistivity simulations with the sample grown at 575 C in Di Salvo 1976, and our Hall effect simulations with the data in the same reference on a sample grown at 600 C.}. The simulation reproduces the low-and high-temperature regimes reasonably well, and importantly includes a peak at 146~K, close to $T_{peak}$ = 165 K in the experimental data. Given the crudeness of the model, and the total neglect of CDW effects, the agreement is remarkable. 

The qualitative understanding of the resistivity within our model depends strongly on thermal population effects. At temperatures below $\sim$100~K, the electron density $n_e=N_d$ is constant, and the resistivity follows the scattering rate only, as in a conventional metal. However, above a characteristic temperature of $\sim$150~K, hole-like carriers become thermally populated to a significant degree (Fig.~\ref{fig:fig2}(a)). As the holes are relatively mobile compared with the electrons, these thermally-activated carriers give a significant contribution to the conductivity. A second effect is that, since $n_e-n_h$ must remains constant, the number of electron-like carriers likewise increases above $\sim$150~K (Fig.~\ref{fig:fig2}(b)). These two effects combine to overcome the near-linearly increasing scattering rate in this regime, yielding a maximum in resistivity at $T_{peak}$. This is followed by a regime of negative $d\rho/dT$ up to room temperature and beyond, as the hole-like carrier density increases at a greater rate than the scattering.

\subsection{Variation with net carrier density}

The net carrier density $N_d$ is an extrinsic and variable parameter of TiSe$_2$ samples. In the standard iodine vapor transport growth method, higher growth temperatures correspond to a larger density of $n-$doping defects \cite{Hildebrand2014}, drastically changing the resistivity \cite{DiSalvo1976,Hildebrand2016}. Post-growth annealing \cite{Huang2017}, deliberate introduction of other transition metal dopants or intercalants \cite{Levy1980,Morosan2006,Luo2016}, and ionic gating \cite{Li2016} have also been used to control the carrier density. To gain some understanding into the variation of the transport properties with this key experimental parameter, in Fig.~\ref{fig:fig3}(a) we consider the effect of varying $N_d$ in our simulations.

The first trend in Fig.~\ref{fig:fig3}(a) is that the lower the carrier density, the greater the height of the peak, with the peak being particularly pronounced for the lowest carrier densities. Within the model, one can understand that the maximum resistivity $\rho_{peak}$ is higher for lower doping levels, since there are simply fewer electrons in the $n_e\approx{}N_d$ regime. However, by room temperature, the thermal population effects dominate, and all moderate dopings give similar behavior in this regime. In this regard, our simulations compare very well with the suppression of the peak height as a function of growth temperature (i.e. extrinsic electron doping) reported by Di Salvo \textit{et al.} \cite{DiSalvo1976} \footnote{Note that in Fig.~\ref{fig:fig3}(a) the impurity scattering rate is held constant for all carrier densities, although experimentally it is likely to scale inversely with $N_e$.}. Campbell \textit{et al.} \cite{CampbellPaglione2018_arxiv} recently argued against any link between the height of the peak and the sample stoichiometry. However, our simulations are in line with the earlier intuition that the two are indeed related \cite{DiSalvo1976,DiSalvo1978,Taguchi1981}. The quantity $\rho{}_{peak}/\rho_{300~K}$ has been sometimes used as an indicator of sample quality \cite{DiSalvo1976,Taguchi1981,Levy1980}. Our simulations show that there is a monotonic inverse dependence of $\rho{}_{peak}/\rho_{300~K}$ on $N_d$, showing that this is indeed a useful indicator of the closeness of the sample to charge compensation. For a substantial doping of $N_d=5\times{}10^{20}$~cm$^{-3}$, corresponding to 0.033 carriers per formula unit, the peak behavior disappears and the resistivity appears generally metallic below room temperature. This is in line with the measured metallic-like resistivity of samples with the highest doping levels \cite{DiSalvo1976,DiSalvo1978,Huang2017,Morosan2006,Hildebrand2016}. Conversely, our model suggests that any sample which has an extrinsic electron doping of less than 0.01 electrons per formula unit will display a prominent peak. 

The second trend in Fig.~\ref{fig:fig3}(a) is that the lower the carrier density, the lower $T_{peak}$. In the model, we find the fewer the number of doped electrons, the fewer thermally activated hole-like carriers are needed to influence the resistivity, and so the peak occurs at a lower temperature. Our simulations may exaggerate this effect somewhat since, as we show later, the CDW transition modulates the exact position of $T_{peak}$, which may tend to dampen the variation found in Fig.~\ref{fig:fig3}(a). Still, the trend is consistent with experimental evidence. For example, the sample grown by sublimation in Di Salvo \textit{et al.} \cite{DiSalvo1976} with $T_{peak}\approx{}$149~K, and the sample grown in Se flux by Campbell \textit{et al.} ($T_{peak}$~= 150~K \cite{CampbellPaglione2018_arxiv}), which are likely to be the most stoichiometric samples (i.e. lowest extrinsic carrier densities), both display a higher $\rho_{peak}$ and a lower $T_{peak}$ compared with samples grown by iodine vapor transport, in which $T_{peak}$ is typically 165~K \cite{DiSalvo1976}.    

\subsection{Robustness of the peak feature}

In addition to the net carrier density, another important parameter in our model is the band gap. We take our value of 74~meV from high-resolution ARPES measurements at 300~K \cite{Watson2019PRL}, but there is considerable disagreement on the magnitude of the normal state band gap in the prior literature \cite{Rossnagel2011}. In Fig.~\ref{fig:fig3}(b) we vary the band gap and find that a peak feature in the resistivity is generically expected, for any relatively narrow band gap. Gap values of 50-100 meV give $T_{peak}$ comparable with the experimental data, correlating with the determination of 74 meV from ARPES \cite{Watson2019PRL}. It is intuitive that $T_{peak}$ scales with the gap, since for a larger gap, higher temperatures are required to activate a significant hole carrier density. 

Furthermore, we show in Fig.~\ref{fig:fig3}(b) that for a small band overlap (negative band gap), no peak appears in our simulations and the resistivity is a monotonically increasing function of temperature, in total disagreement with the experimental resistivity. The simulations thus support the simple intuition that the negative $d\rho{}/dT$ around room temperature is indicative of a narrow band gap in the normal state. While a semimetallic scenario for the normal state was proposed by Velebit \textit{et al.}\cite{Velebit2016PRB}, this could only be reconciled with the resistivity data in the normal phase by assuming a highly anomalous temperature-dependence of the scattering rate (i.e. mobilities increasing with T at all temperatures above $T_c$). In contrast, here we show that a narrow band gap, with a conventional scattering rate that increases monotonically with temperature, qualitatively reproduces the resistivity across the whole temperature range quite well, as well as showing similar trends with variation of $N_d$ to the experiments. Thus on the basis of the successful simulations in this paper, as well as the evidence from photoemission \cite{Watson2019PRL}, we would argue that a narrow band gap and a conventional temperature-dependence of the scattering rates (apart from, perhaps, at temperatures very close to $T_c$) is the scenario that best accounts for the normal state electronic structure and transport properties of TiSe$_2$. 

To further test the robustness of the peak feature in resistivity, in Fig.~\ref{fig:fig3}(c) we vary the electron effective mass. So far, we have used an effective mass which matches the electron band dispersion in the L-A direction (Fig.~\ref{fig:fig1}(a)). This is likely an over-estimate, however, since the band is anisotropic and the effective mass is expected to be lighter in the L-M and L-H directions. Fig~\ref{fig:fig3}(c) shows that the peak is slightly more pronounced for heavier electrons, in which case the thermally-activated holes are relatively more mobile. However qualitatively, the appearance of a peak in resistivity is robust under even substantial variation of this parameter. This remains generally true under moderate variations in all other parameters, so long as the holes are substantially more mobile, the band gap is positive, and the extrinsic electron-like carrier density is not too large.    

\begin{figure}[b]
	\centering
	\includegraphics[width=\linewidth]{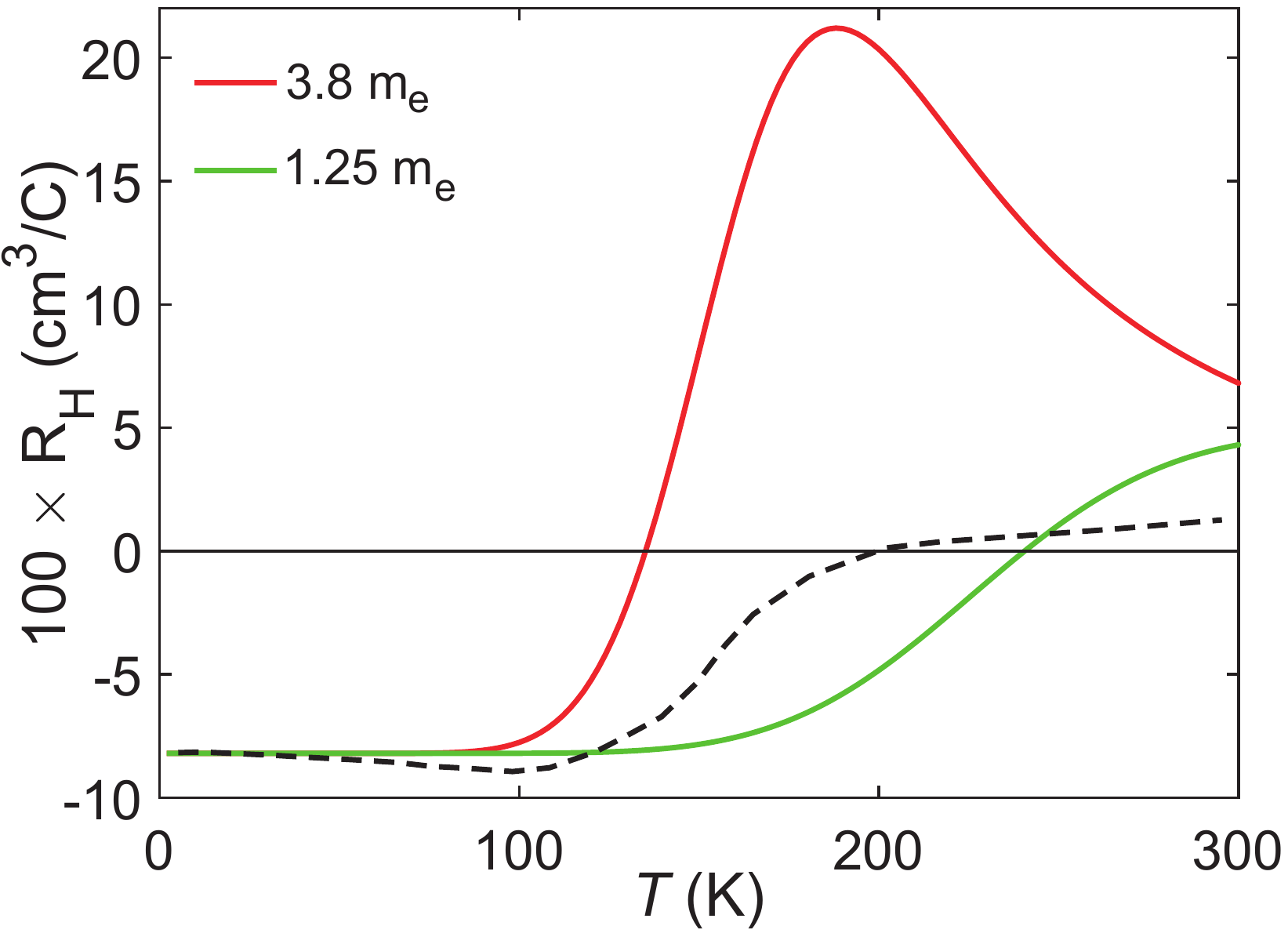}
	\caption{Simulation of the Hall coefficient, for two values of the electron effective mass, compared with experimental data \cite{DiSalvo1976} (dashed line). The saturation of $R_H$ below $\sim$100~K and a sign-change of $R_H$ by room temperature are qualitatively reproduced.}
	\label{fig:fig4}
\end{figure}

\subsection{Sign-change of the Hall coefficient}

Our model also gives insight into the Hall effect in TiSe$_2$. A long-standing problem has been why the samples, known to be $n$-doped, show a crossover from a negative Hall coefficient at low temperatures to a positive value at room temperature \cite{DiSalvo1976,Taguchi1981,Wilson1978}. As shown in Fig.~\ref{fig:fig4}, we also find a sign-change of the Hall effect in our model without CDW order, where we use the standard two-carrier transport equations to calculate the Hall response:

\begin{equation}
R_H = \frac{\mu_h^2n_h-\mu_e^2n_e}{|e|(\mu_hn_h+mu_en_e)^2}
\end{equation}

Once again, the relevant physics is the freezing-out of hole like carriers; below $\sim$100~K, there are no thermally activated holes, and the Hall effect reduces to the simple one carrier model, $R_H\approx-1/N_d|e|$. However, as the holes become activated, due to their relatively high mobility, they quickly become very relevant in the Hall effect. By room temperature, even though $n_h<n_e$ due to the extrinsic electron doping, the Hall coefficient is positive due to the higher hole mobility. Similar logic can likely account for the positive Seebeck coefficient at high temperatures \cite{DiSalvo1976}. We note that the Hall effect is particularly sensitive to the relative mobilities of electrons and holes, and here our assumption that the holes and electrons have the same scattering rate may cause a greater discrepancy with the experimental data, compared with the resistivity simulations. In Fig.~\ref{fig:fig4} we also show a simulation with a lower $m^*_e$, which gives better agreement at higher temperatures, reflecting the fact that the ratio of mobilities in our simple model may not be optimal. In addition, using the simple two-carrier equations may not completely encode the fact that there is one hole and three electron bands in the model. Nonetheless, the qualitative agreement in the crossover behavior again adds validity to our simplified model approach, and points again to the key role played by thermally-activated carriers.

The sign change of the Hall effect was reported to occur at 181~K \cite{Taguchi1981}. This proximity to $T_{CDW}$ might naively suggest a connection with changes in the Fermi surface at $T_{CDW}$. However, our simulations show that a sign-change of the Hall effect, somewhere between 100 and 300 K, is generically to be expected simply from thermal activation of hole-like carriers. A sign-change of the Hall coefficient, or any slope-change, therefore is not a good metric of $T_{CDW}$ \cite{Li2016}, due to the dominance of thermal population effects in the measurements.

\begin{figure*}
	\centering
	\includegraphics[width=\linewidth]{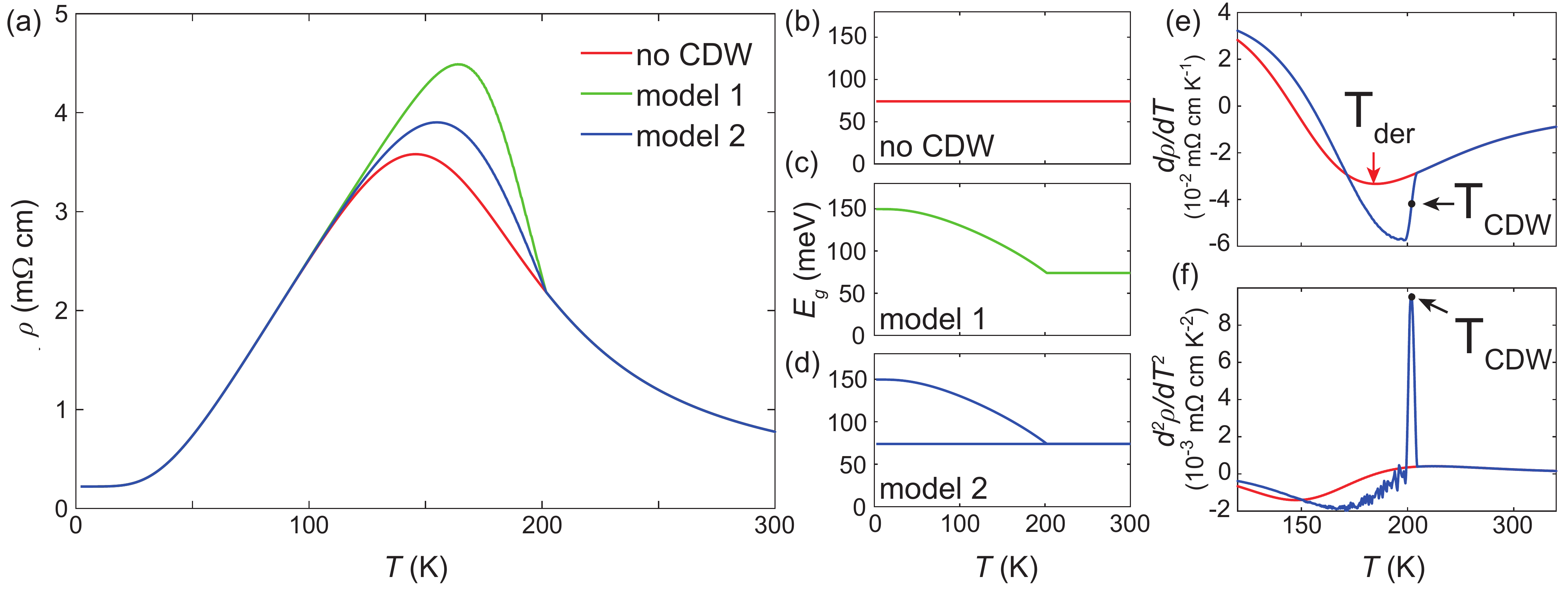}
	\caption{(a) Resistivity curves, in which different approaches to the CDW are implemented below $T_{CDW}$=202~K. (b-d) Corresponding temperature-dependence of the band gap in each case; in model 1 the band gap fully opens up, while in model 2 the hole DOS is split into two components below $T_{CDW}$, one of which doesn't evolve with temperature. (e) Derivative of resistivity with respect to temperature, for model 2 and without CDW. Even without any CDW, $d\rho/dT$ shows a minimum at $T_{der}$, which in this case is meaningless. (f) A sharp peak in the second derivative is the better indicator for $T_{CDW}$.}
	\label{fig:fig5}
\end{figure*}

\subsection{Effects of CDW order on resistivity}

Having shown that the resistivity can be reasonably understood even without any CDW order, we now consider what effects the CDW could have on the transport properties. The CDW of TiSe$_2$ significantly reconstructs the band structure, changes the effective masses of both the holes and the electron bands, will have an impact on scattering rates, and more \cite{Watson2019PRL,Rossnagel2011}. Given the crude level of our model, it is not realistic to attempt to simulate all of these effects, but we can gain some intuition on the possible effects of the CDW order with two simplified models. At $T_{CDW}$, the hole bands hybridize across the band gap with the conduction bands. In the simplest case, therefore, one might consider simply modifying the band gap as $E_g(T)  = \sqrt{E_{g0}^2+\Delta{}(T)^2}$, where $\Delta{}(T)$ takes an order-parameter-like form; we choose $\Delta{}(T)=\Delta_0\tanh{(1.16\sqrt{\frac{T_{CDW}}{T}-1})}$ following Ref.~ \cite{Chen2015}. This is encoded by our model 1 in Fig.~\ref{fig:fig5}, which is similar in spirit to the approach of Ref.~\cite{Monney2009PhysicaB}. At $T_{CDW}$, there is a change of slope below which resistance rises more rapidly than for the non-CDW case. Despite the large energy scale of the coupling ($\Delta_0$~= 130~meV), however, the effects on the resistivity are not particularly dramatic: model 1 only moderately increases $T_{peak}$ and the absolute magnitude of the resistivity,  while the overall shape remains similar. 

Model 1 in fact overestimates the effects of the CDW. In fact, when the CDW occurs, the hybridization is $k_z$-selective; the hole bands at $\Gamma$ hybridize strongly and get shifted to higher binding energies, while the hole bands at A are only weakly affected \cite{Watson2019PRL}. This idea is implemented in model 2, where we shift half of the density of states of the hole band below $T_{CDW}$ according to the increased gap of model 1, but leave the other half at an unchanged energy. As one might expect, this has an even more muted effect on the total resistivity, with only a weak change of slope at $T_{CDW}$. This subtle deviation of the resistivity below $T_{CDW}$ is consistent with the experimental changes in resistivity through $T_{CDW}$, where at most a modest change of slope is detected.

Neither of these simple models is even close to accounting for all of the CDW physics, and in particular we do not consider any modifications to the electron bands through the transition. In fact, experimentally it is found that the electron bands substantially change shape through $T_{CDW}$, and the conduction band minimum sinks to higher binding energies, causing the band gap to actually be smaller in the CDW phase \cite{Watson2019PRL}. However, the most physically relevant parameter is not the band gap \textit{per se}, but the binding energy of the hole-like states, and especially in the case of model 2, these are reasonably consistent with the ARPES measurements. Thus, in the spirit of emphasizing the temperature-dependence of thermally-populated hole-like carriers, one can come to a qualitative understanding. Upon cooling the sample, the CDW accelerates the freezing-out of hole-like carriers as they are shifted to higher binding energy, increasing resistance below $T_{CDW}$ and bringing $T_{peak}$ to slightly higher temperatures; for example, model 2 increases $T_{peak}$ from 146 to 154~K. However, this freezing out of hole-like carriers would have happened even without the CDW order. We would therefore argue that, although the peak in resistivity may be somewhat modulated by the CDW ordering, this is only a secondary effect. Fundamentally, there is no causal link between the peak in resistivity and the CDW, and the peak should primarily be understood it in terms of the thermal population effects described earlier. 

\subsection{Criteria for the determination of CDW temperature}

As in the simulations, a change of slope in resistivity measurements is a mark of the onset of CDW order, but in practice a variety of procedures have been used in the literature to extract $T_c$ from experimental data. While the comparison of such criteria may seem a little mundane, the reliable determination of $T_c$ becomes crucial for the construction of phase diagrams based on tuning the properties of TiSe$_2$. Superconductivity has been found in TiSe$_2$ close to the critical point for the suppression of long-range CDW order via pressure \cite{Kusmartseva2009}, charge carrier doping \cite{Li2016}, or intercalation \cite{Morosan2006,Morosan2010PRB,Luo2016}. However, for a full understanding of the interplay of CDW and superconducting orders, it is important to be confident of the determination of the CDW phase boundary, which in some studies has been partially or solely delineated by the analysis of transport measurements. Thus there is a need to firmly establish the appropriate criterion for the determination of $T_c$ from such measurements.

One procedure used in the literature is to take the first derivative of the data, and then to use the minimum of $d\rho{}/dT$ (maximum of $-d\rho{}/dT$) as a criterion for $T_{CDW}$ \cite{Wegner2018_arxiv,Castellan2013,Li2016APL}. However, our simulations without CDW order also give a prominent minimum in $d\rho{}/dT$ (Fig.~\ref{fig:fig5}(e)), even though nothing is physically happening at this temperature. Moreover, in the simulations including CDW order, the minimum of $d\rho{}/dT$ does not coincide with $T_{CDW}$. The minimum of $d\rho{}/dT$ is therefore not an appropriate criterion. A more physical approach is to look for a discontinuity in $d\rho{}/dT$, since this should jump across $T_{CDW}$ \cite{DiSalvo1976}. While our simulations support this idea in principle, in practice it may be difficult to precisely identify the midpoint of the step-like feature in the first derivative, given there is a substantially varying background. Taguchi \textit{et al.} \cite{Taguchi1981}, and more recently
Campbell \textit{et al.} \cite{CampbellPaglione2018_arxiv}, argued that looking for a small but sharp peak in the second derivative is a better criterion for determining $T_{CDW}$. In agreement with this, our analysis presented Fig. 5(f) shows a sharp peak centred at $T_c$ in the second derivative, although we note that discerning this peak in practice may often be complicated by the amplification of experimental noise in such second derivative analysis. Identifying the CDW onset temperature in TiSe$_2$ may, in general, therefore require additional measurements such as specific heat \cite{Craven1978}, or the observation of superlattice reflections in scattering measurements \cite{Kogar2017PRL}.

\subsection{Discussion}

In our simulations, the resistivity in the low temperature limit is always metallic-like ($d\rho{}/dT>0$). For experimental samples grown by iodine vapor transport this is also usually the case, with the electron carriers coming from a finite extrinsic carrier density $N_d$. Experimentally, this arises from crystal imperfections: Se vacancies, intercalation of excess Ti, and residual iodine inclusions \cite{Hildebrand2014}. Correspondingly, a coherent electron-like Fermi surface is observed by ARPES at low temperatures \cite{Watson2019PRL}, and there is a small but finite Sommerfeld term in specific heat \cite{Craven1978}. Such samples typically show an approximately constant Hall coefficient below 100~K \cite{DiSalvo1976}, indicating that $N_d$ does not change significantly down to low temperature. 

However, different methods of extrinsic doping lead to strongly varying results. For example Ti$_{1-x}$Ta$_x$Se$_2$ shows metallic-like behavior with $d\rho{}/dT>0$ and a superconducting dome, but Ti$_{1-x}$Nb$_x$Se$_2$ gives insulating-like $d\rho{}/dT<0$ all the way to room temperature \cite{Levy1980,Luo2016}, even though the Nb and Ta dopants are nominally isoelectronic. The details of how the dopant species interacts with the host matrix are thus highly relevant, as Nb dopants appear to cause substantially more disorder than Ta \cite{Luo2016}. In other cases, such as lightly-doped Pd$_x$TiSe$_2$ \cite{Morosan2010PRB} and some of the pressure-grown TiSe$_2$ samples reported by Campbell \textit{et al.} \cite{CampbellPaglione2018_arxiv}, the resistivity shows similar behavior to the iodine vapor transport-grown samples, with a slope-change at $T_c$ and a broad peak, but then exhibits a steep upturn in resistivity below 80-100~K, while the Hall coefficient only saturates at a much lower temperature, and at a larger magnitude \cite{CampbellPaglione2018_arxiv}. This may reflect a significant fraction of the donated electrons becoming localized at low temperatures. We note that $N_d$ is treated as a constant in our simulations, and so no such upturn of the resistivity is found. To our knowledge, there has been no successful synthesis of $p$-type TiSe$_2$: all reported samples have a negative Hall coefficient at low temperatures, despite efforts by Levy \textit{et al.} to introduce acceptors \cite{Levy1980}. We predict that any such $p$-type samples could show drastically different resistivity properties, and it is an open question whether the CDW would also persist in such samples.     

Much of the more recent literature on TiSe$_2$ has addressed the question of whether it is a semiconductor or semimetal, especially in the normal state \cite{Rossnagel2002,Rossnagel2011,Li2007,Monney2010PRB,May2011PRL,Velebit2016PRB,Hellgren2017}. Here, we have shown that the presence of a narrow band gap can naturally account for the overall form of the resistivity including the broad peak. In combination with recent high-resolution photoemission \cite{Watson2019PRL}, this provides compelling evidence in favor of a narrow band gap scenario. 

In conclusion, we have shown that a minimal model based on the normal-state electronic structure of TiSe$_2$ can qualitatively reproduce its known temperature-dependent transport behavior. Starting from the assumption of a narrow band gap, a crucial component is the temperature-dependence of the hole and electron carrier densities. Without incorporating any CDW-related changes, the broad peak in resistivity can be reproduced, and a sign-change in the Hall effect is also found. Thus neither of these experimental signatures is directly related to the CDW. We included CDW order into the model with crude approximations, and showed that, despite large energy scales being involved, the effect on resistivity was relatively muted, qualitatively accounting for the typically weak signatures observed in transport at $T_c$. While the CDW in TiSe$_2$ remains a highly interesting problem, our simple but robust simulations show that there is another effect which is more relevant for the overall understanding of the transport data of TiSe$_2$, which is the thermal activation of mobile hole-like carriers. 

\section{Acknowledgments}
We thank K.~Rossnagel for useful discussions, and all the authors of Ref.~\cite{Watson2019PRL} who contributed to the photoemission data shown in Fig.~1. We gratefully acknowledge support from The Leverhulme Trust (Grant No. RL-2016-006) and The Royal Society. A.B. acknowledges support from The Physics Trust and the Student Staff Council of the School of Physics and Astronomy, University of St Andrews.


\begin{thebibliography}{34}%
	\makeatletter
	\providecommand \@ifxundefined [1]{%
		\@ifx{#1\undefined}
	}%
	\providecommand \@ifnum [1]{%
		\ifnum #1\expandafter \@firstoftwo
		\else \expandafter \@secondoftwo
		\fi
	}%
	\providecommand \@ifx [1]{%
		\ifx #1\expandafter \@firstoftwo
		\else \expandafter \@secondoftwo
		\fi
	}%
	\providecommand \natexlab [1]{#1}%
	\providecommand \enquote  [1]{``#1''}%
	\providecommand \bibnamefont  [1]{#1}%
	\providecommand \bibfnamefont [1]{#1}%
	\providecommand \citenamefont [1]{#1}%
	\providecommand \href@noop [0]{\@secondoftwo}%
	\providecommand \href [0]{\begingroup \@sanitize@url \@href}%
	\providecommand \@href[1]{\@@startlink{#1}\@@href}%
	\providecommand \@@href[1]{\endgroup#1\@@endlink}%
	\providecommand \@sanitize@url [0]{\catcode `\\12\catcode `\$12\catcode
		`\&12\catcode `\#12\catcode `\^12\catcode `\_12\catcode `\%12\relax}%
	\providecommand \@@startlink[1]{}%
	\providecommand \@@endlink[0]{}%
	\providecommand \url  [0]{\begingroup\@sanitize@url \@url }%
	\providecommand \@url [1]{\endgroup\@href {#1}{\urlprefix }}%
	\providecommand \urlprefix  [0]{URL }%
	\providecommand \Eprint [0]{\href }%
	\providecommand \doibase [0]{http://dx.doi.org/}%
	\providecommand \selectlanguage [0]{\@gobble}%
	\providecommand \bibinfo  [0]{\@secondoftwo}%
	\providecommand \bibfield  [0]{\@secondoftwo}%
	\providecommand \translation [1]{[#1]}%
	\providecommand \BibitemOpen [0]{}%
	\providecommand \bibitemStop [0]{}%
	\providecommand \bibitemNoStop [0]{.\EOS\space}%
	\providecommand \EOS [0]{\spacefactor3000\relax}%
	\providecommand \BibitemShut  [1]{\csname bibitem#1\endcsname}%
	\let\auto@bib@innerbib\@empty
	\bibitem [{\citenamefont {Di~Salvo}\ \emph {et~al.}(1976)\citenamefont
		{Di~Salvo}, \citenamefont {Moncton},\ and\ \citenamefont
		{Waszczak}}]{DiSalvo1976}%
	\BibitemOpen
	\bibfield  {author} {\bibinfo {author} {\bibfnamefont {F.~J.}\ \bibnamefont
			{Di~Salvo}}, \bibinfo {author} {\bibfnamefont {D.~E.}\ \bibnamefont
			{Moncton}}, \ and\ \bibinfo {author} {\bibfnamefont {J.~V.}\ \bibnamefont
			{Waszczak}},\ }\bibfield  {title} {\enquote {\bibinfo {title} {{Electronic
					properties and superlattice formation in the semimetal
					${\mathrm{TiSe}}_{2}$}},}\ }\href {\doibase 10.1103/PhysRevB.14.4321}
	{\bibfield  {journal} {\bibinfo  {journal} {Phys. Rev. B}\ }\textbf {\bibinfo
			{volume} {14}},\ \bibinfo {pages} {4321--4328} (\bibinfo {year}
		{1976})}\BibitemShut {NoStop}%
	\bibitem [{\citenamefont {Craven}\ \emph {et~al.}(1978)\citenamefont {Craven},
		\citenamefont {Salvo},\ and\ \citenamefont {Hsu}}]{Craven1978}%
	\BibitemOpen
	\bibfield  {author} {\bibinfo {author} {\bibfnamefont {R.A.}\ \bibnamefont
			{Craven}}, \bibinfo {author} {\bibfnamefont {F.J.~Di}\ \bibnamefont {Salvo}},
		\ and\ \bibinfo {author} {\bibfnamefont {F.S.L.}\ \bibnamefont {Hsu}},\
	}\bibfield  {title} {\enquote {\bibinfo {title} {{Mechanisms for the 200 K
					transition in ${\mathrm{TiSe}}_{2}$: A measurement of the specific heat}},}\
	}\href {\doibase https://doi.org/10.1016/0038-1098(78)91165-1} {\bibfield
		{journal} {\bibinfo  {journal} {Solid State Communications}\ }\textbf
		{\bibinfo {volume} {25}},\ \bibinfo {pages} {39 -- 42} (\bibinfo {year}
		{1978})}\BibitemShut {NoStop}%
	\bibitem [{\citenamefont {Taguchi}\ \emph {et~al.}(1981)\citenamefont
		{Taguchi}, \citenamefont {Asai}, \citenamefont {Watanabe},\ and\
		\citenamefont {Oka}}]{Taguchi1981}%
	\BibitemOpen
	\bibfield  {author} {\bibinfo {author} {\bibfnamefont {I.}~\bibnamefont
			{Taguchi}}, \bibinfo {author} {\bibfnamefont {M.}~\bibnamefont {Asai}},
		\bibinfo {author} {\bibfnamefont {Y.}~\bibnamefont {Watanabe}}, \ and\
		\bibinfo {author} {\bibfnamefont {M.}~\bibnamefont {Oka}},\ }\bibfield
	{title} {\enquote {\bibinfo {title} {{Transport properties of iodine-free
					TiSe$_2$}},}\ }\href {\doibase https://doi.org/10.1016/0378-4363(81)90234-5}
	{\bibfield  {journal} {\bibinfo  {journal} {Physica B+C}\ }\textbf {\bibinfo
			{volume} {105}},\ \bibinfo {pages} {146 -- 150} (\bibinfo {year}
		{1981})}\BibitemShut {NoStop}%
	\bibitem [{\citenamefont {Rossnagel}(2011)}]{Rossnagel2011}%
	\BibitemOpen
	\bibfield  {author} {\bibinfo {author} {\bibfnamefont {K}~\bibnamefont
			{Rossnagel}},\ }\bibfield  {title} {\enquote {\bibinfo {title} {{On the
					origin of charge-density waves in select layered transition-metal
					dichalcogenides}},}\ }\href
	{http://stacks.iop.org/0953-8984/23/i=21/a=213001} {\bibfield  {journal}
		{\bibinfo  {journal} {Journal of Physics: Condensed Matter}\ }\textbf
		{\bibinfo {volume} {23}},\ \bibinfo {pages} {213001} (\bibinfo {year}
		{2011})}\BibitemShut {NoStop}%
	\bibitem [{\citenamefont {Di~Salvo}\ and\ \citenamefont
		{Waszczak}(1978)}]{DiSalvo1978}%
	\BibitemOpen
	\bibfield  {author} {\bibinfo {author} {\bibfnamefont {F.~J.}\ \bibnamefont
			{Di~Salvo}}\ and\ \bibinfo {author} {\bibfnamefont {J.~V.}\ \bibnamefont
			{Waszczak}},\ }\bibfield  {title} {\enquote {\bibinfo {title} {{Transport
					properties and the phase transition in
					${\mathrm{Ti}}_{1\ensuremath{-}x}{M}_{x}{\mathrm{Se}}_{2}
					(M=\mathrm{Ta}~\mathrm{or}~\mathrm{V})$}},}\ }\href {\doibase
		10.1103/PhysRevB.17.3801} {\bibfield  {journal} {\bibinfo  {journal} {Phys.
				Rev. B}\ }\textbf {\bibinfo {volume} {17}},\ \bibinfo {pages} {3801--3807}
		(\bibinfo {year} {1978})}\BibitemShut {NoStop}%
	\bibitem [{\citenamefont {Levy}(1980)}]{Levy1980}%
	\BibitemOpen
	\bibfield  {author} {\bibinfo {author} {\bibfnamefont {F}~\bibnamefont
			{Levy}},\ }\bibfield  {title} {\enquote {\bibinfo {title} {{The influence of
					impurities on the electrical properties of TiSe$_2$ single crystals}},}\
	}\href {http://stacks.iop.org/0022-3719/13/i=15/a=014} {\bibfield  {journal}
		{\bibinfo  {journal} {Journal of Physics C: Solid State Physics}\ }\textbf
		{\bibinfo {volume} {13}},\ \bibinfo {pages} {2901} (\bibinfo {year}
		{1980})}\BibitemShut {NoStop}%
	\bibitem [{\citenamefont {Monney}\ \emph {et~al.}(2009)\citenamefont {Monney},
		\citenamefont {Cercellier}, \citenamefont {Battaglia}, \citenamefont
		{Schwier}, \citenamefont {Didiot}, \citenamefont {Garnier}, \citenamefont
		{Beck},\ and\ \citenamefont {Aebi}}]{Monney2009PhysicaB}%
	\BibitemOpen
	\bibfield  {author} {\bibinfo {author} {\bibfnamefont {C.}~\bibnamefont
			{Monney}}, \bibinfo {author} {\bibfnamefont {H.}~\bibnamefont {Cercellier}},
		\bibinfo {author} {\bibfnamefont {C.}~\bibnamefont {Battaglia}}, \bibinfo
		{author} {\bibfnamefont {E.F.}\ \bibnamefont {Schwier}}, \bibinfo {author}
		{\bibfnamefont {C.}~\bibnamefont {Didiot}}, \bibinfo {author} {\bibfnamefont
			{M.G.}\ \bibnamefont {Garnier}}, \bibinfo {author} {\bibfnamefont
			{H.}~\bibnamefont {Beck}}, \ and\ \bibinfo {author} {\bibfnamefont
			{P.}~\bibnamefont {Aebi}},\ }\bibfield  {title} {\enquote {\bibinfo {title}
			{{Temperature dependence of the excitonic insulator phase model in
					1T-TiSe$_2$}},}\ }\href {\doibase
		https://doi.org/10.1016/j.physb.2009.07.047} {\bibfield  {journal} {\bibinfo
			{journal} {Physica B: Condensed Matter}\ }\textbf {\bibinfo {volume} {404}},\
		\bibinfo {pages} {3172 -- 3175} (\bibinfo {year} {2009})}\BibitemShut
	{NoStop}%
	\bibitem [{\citenamefont {Monney}\ \emph
		{et~al.}(2010{\natexlab{a}})\citenamefont {Monney}, \citenamefont {Schwier},
		\citenamefont {Garnier}, \citenamefont {Mariotti}, \citenamefont {Didiot},
		\citenamefont {Beck}, \citenamefont {Aebi}, \citenamefont {Cercellier},
		\citenamefont {Marcus}, \citenamefont {Battaglia}, \citenamefont {Berger},\
		and\ \citenamefont {Titov}}]{Monney2010PRB}%
	\BibitemOpen
	\bibfield  {author} {\bibinfo {author} {\bibfnamefont {C.}~\bibnamefont
			{Monney}}, \bibinfo {author} {\bibfnamefont {E.~F.}\ \bibnamefont {Schwier}},
		\bibinfo {author} {\bibfnamefont {M.~G.}\ \bibnamefont {Garnier}}, \bibinfo
		{author} {\bibfnamefont {N.}~\bibnamefont {Mariotti}}, \bibinfo {author}
		{\bibfnamefont {C.}~\bibnamefont {Didiot}}, \bibinfo {author} {\bibfnamefont
			{H.}~\bibnamefont {Beck}}, \bibinfo {author} {\bibfnamefont {P.}~\bibnamefont
			{Aebi}}, \bibinfo {author} {\bibfnamefont {H.}~\bibnamefont {Cercellier}},
		\bibinfo {author} {\bibfnamefont {J.}~\bibnamefont {Marcus}}, \bibinfo
		{author} {\bibfnamefont {C.}~\bibnamefont {Battaglia}}, \bibinfo {author}
		{\bibfnamefont {H.}~\bibnamefont {Berger}}, \ and\ \bibinfo {author}
		{\bibfnamefont {A.~N.}\ \bibnamefont {Titov}},\ }\bibfield  {title} {\enquote
		{\bibinfo {title} {{Temperature-dependent photoemission on
					$1T{\text{-TiSe}}_{2}$: Interpretation within the exciton condensate phase
					model}},}\ }\href {\doibase 10.1103/PhysRevB.81.155104} {\bibfield  {journal}
		{\bibinfo  {journal} {Phys. Rev. B}\ }\textbf {\bibinfo {volume} {81}},\
		\bibinfo {pages} {155104} (\bibinfo {year} {2010}{\natexlab{a}})}\BibitemShut
	{NoStop}%
	\bibitem [{\citenamefont {Monney}\ \emph
		{et~al.}(2010{\natexlab{b}})\citenamefont {Monney}, \citenamefont {Schwier},
		\citenamefont {Garnier}, \citenamefont {Mariotti}, \citenamefont {Didiot},
		\citenamefont {Cercellier}, \citenamefont {Marcus}, \citenamefont {Berger},
		\citenamefont {Titov}, \citenamefont {Beck},\ and\ \citenamefont
		{Aebi}}]{Monney2010NJP}%
	\BibitemOpen
	\bibfield  {author} {\bibinfo {author} {\bibfnamefont {C}~\bibnamefont
			{Monney}}, \bibinfo {author} {\bibfnamefont {E~F}\ \bibnamefont {Schwier}},
		\bibinfo {author} {\bibfnamefont {M~G}\ \bibnamefont {Garnier}}, \bibinfo
		{author} {\bibfnamefont {N}~\bibnamefont {Mariotti}}, \bibinfo {author}
		{\bibfnamefont {C}~\bibnamefont {Didiot}}, \bibinfo {author} {\bibfnamefont
			{H}~\bibnamefont {Cercellier}}, \bibinfo {author} {\bibfnamefont
			{J}~\bibnamefont {Marcus}}, \bibinfo {author} {\bibfnamefont {H}~\bibnamefont
			{Berger}}, \bibinfo {author} {\bibfnamefont {A~N}\ \bibnamefont {Titov}},
		\bibinfo {author} {\bibfnamefont {H}~\bibnamefont {Beck}}, \ and\ \bibinfo
		{author} {\bibfnamefont {P}~\bibnamefont {Aebi}},\ }\bibfield  {title}
	{\enquote {\bibinfo {title} {{Probing the exciton condensate phase in
					1T-{TiSe}$_2$ with photoemission}},}\ }\href {\doibase
		10.1088/1367-2630/12/12/125019} {\bibfield  {journal} {\bibinfo  {journal}
			{New Journal of Physics}\ }\textbf {\bibinfo {volume} {12}},\ \bibinfo
		{pages} {125019} (\bibinfo {year} {2010}{\natexlab{b}})}\BibitemShut
	{NoStop}%
	\bibitem [{\citenamefont {Koley}\ \emph {et~al.}(2014)\citenamefont {Koley},
		\citenamefont {Laad}, \citenamefont {Vidhyadhiraja},\ and\ \citenamefont
		{Taraphder}}]{Koley2014}%
	\BibitemOpen
	\bibfield  {author} {\bibinfo {author} {\bibfnamefont {S.}~\bibnamefont
			{Koley}}, \bibinfo {author} {\bibfnamefont {M.~S.}\ \bibnamefont {Laad}},
		\bibinfo {author} {\bibfnamefont {N.~S.}\ \bibnamefont {Vidhyadhiraja}}, \
		and\ \bibinfo {author} {\bibfnamefont {A.}~\bibnamefont {Taraphder}},\
	}\bibfield  {title} {\enquote {\bibinfo {title} {{Preformed excitons, orbital
					selectivity, and charge density wave order in
					$1T\text{\ensuremath{-}}{\mathrm{TiSe}}_{2}$}},}\ }\href {\doibase
		10.1103/PhysRevB.90.115146} {\bibfield  {journal} {\bibinfo  {journal} {Phys.
				Rev. B}\ }\textbf {\bibinfo {volume} {90}},\ \bibinfo {pages} {115146}
		(\bibinfo {year} {2014})}\BibitemShut {NoStop}%
	\bibitem [{\citenamefont {Hildebrand}\ \emph {et~al.}(2016)\citenamefont
		{Hildebrand}, \citenamefont {Jaouen}, \citenamefont {Didiot}, \citenamefont
		{Razzoli}, \citenamefont {Monney}, \citenamefont {Mottas}, \citenamefont
		{Ubaldini}, \citenamefont {Berger}, \citenamefont {Barreteau}, \citenamefont
		{Beck}, \citenamefont {Bowler},\ and\ \citenamefont {Aebi}}]{Hildebrand2016}%
	\BibitemOpen
	\bibfield  {author} {\bibinfo {author} {\bibfnamefont {B.}~\bibnamefont
			{Hildebrand}}, \bibinfo {author} {\bibfnamefont {T.}~\bibnamefont {Jaouen}},
		\bibinfo {author} {\bibfnamefont {C.}~\bibnamefont {Didiot}}, \bibinfo
		{author} {\bibfnamefont {E.}~\bibnamefont {Razzoli}}, \bibinfo {author}
		{\bibfnamefont {G.}~\bibnamefont {Monney}}, \bibinfo {author} {\bibfnamefont
			{M.-L.}\ \bibnamefont {Mottas}}, \bibinfo {author} {\bibfnamefont
			{A.}~\bibnamefont {Ubaldini}}, \bibinfo {author} {\bibfnamefont
			{H.}~\bibnamefont {Berger}}, \bibinfo {author} {\bibfnamefont
			{C.}~\bibnamefont {Barreteau}}, \bibinfo {author} {\bibfnamefont
			{H.}~\bibnamefont {Beck}}, \bibinfo {author} {\bibfnamefont {D.~R.}\
			\bibnamefont {Bowler}}, \ and\ \bibinfo {author} {\bibfnamefont
			{P.}~\bibnamefont {Aebi}},\ }\bibfield  {title} {\enquote {\bibinfo {title}
			{{Short-range phase coherence and origin of the
					$1T\ensuremath{-}{\mathrm{TiSe}}_{2}$ charge density wave}},}\ }\href
	{\doibase 10.1103/PhysRevB.93.125140} {\bibfield  {journal} {\bibinfo
			{journal} {Phys. Rev. B}\ }\textbf {\bibinfo {volume} {93}},\ \bibinfo
		{pages} {125140} (\bibinfo {year} {2016})}\BibitemShut {NoStop}%
	\bibitem [{\citenamefont {Wilson}(1978)}]{Wilson1978}%
	\BibitemOpen
	\bibfield  {author} {\bibinfo {author} {\bibfnamefont {J.~A.}\ \bibnamefont
			{Wilson}},\ }\bibfield  {title} {\enquote {\bibinfo {title} {{Modelling the
					contrasting semimetallic characters of TiS$_2$ and TiSe$_2$}},}\ }\href
	{\doibase 10.1002/pssb.2220860102} {\bibfield  {journal} {\bibinfo  {journal}
			{physica status solidi (b)}\ }\textbf {\bibinfo {volume} {86}},\ \bibinfo
		{pages} {11--36} (\bibinfo {year} {1978})}\BibitemShut {NoStop}%
	\bibitem [{\citenamefont {Rossnagel}\ \emph {et~al.}(2002)\citenamefont
		{Rossnagel}, \citenamefont {Kipp},\ and\ \citenamefont
		{Skibowski}}]{Rossnagel2002}%
	\BibitemOpen
	\bibfield  {author} {\bibinfo {author} {\bibfnamefont {K.}~\bibnamefont
			{Rossnagel}}, \bibinfo {author} {\bibfnamefont {L.}~\bibnamefont {Kipp}}, \
		and\ \bibinfo {author} {\bibfnamefont {M.}~\bibnamefont {Skibowski}},\
	}\bibfield  {title} {\enquote {\bibinfo {title} {{Charge-density-wave phase
					transition in $1T\ensuremath{-}{\mathrm{TiSe}}_{2}:$ Excitonic insulator
					versus band-type Jahn-Teller mechanism}},}\ }\href {\doibase
		10.1103/PhysRevB.65.235101} {\bibfield  {journal} {\bibinfo  {journal} {Phys.
				Rev. B}\ }\textbf {\bibinfo {volume} {65}},\ \bibinfo {pages} {235101}
		(\bibinfo {year} {2002})}\BibitemShut {NoStop}%
	\bibitem [{\citenamefont {{Campbell}}\ \emph {et~al.}(2018)\citenamefont
		{{Campbell}}, \citenamefont {{Eckberg}}, \citenamefont {{Zavalij}},\ and\
		\citenamefont {{Paglione}}}]{CampbellPaglione2018_arxiv}%
	\BibitemOpen
	\bibfield  {author} {\bibinfo {author} {\bibfnamefont {D.~J.}\ \bibnamefont
			{{Campbell}}}, \bibinfo {author} {\bibfnamefont {C.}~\bibnamefont
			{{Eckberg}}}, \bibinfo {author} {\bibfnamefont {P.~Y.}\ \bibnamefont
			{{Zavalij}}}, \ and\ \bibinfo {author} {\bibfnamefont {J.}~\bibnamefont
			{{Paglione}}},\ }\bibfield  {title} {\enquote {\bibinfo {title} {{Intrinsic
					Insulating Ground State in Transition Metal Dichalcogenide TiSe$_2$}},}\
	}\href {https://arxiv.org/abs/1809.09467} {\bibfield  {journal} {\bibinfo
			{journal} {arXiv:1809.09467}\ } (\bibinfo {year} {2018})}\BibitemShut
	{NoStop}%
	\bibitem [{\citenamefont {Watson}\ \emph {et~al.}(2019)\citenamefont {Watson},
		\citenamefont {Clark}, \citenamefont {Mazzola}, \citenamefont
		{Markovi\ifmmode~\acute{c}\else \'{c}\fi{}}, \citenamefont {Sunko},
		\citenamefont {Kim}, \citenamefont {Rossnagel},\ and\ \citenamefont
		{King}}]{Watson2019PRL}%
	\BibitemOpen
	\bibfield  {author} {\bibinfo {author} {\bibfnamefont {Matthew~D.}\
			\bibnamefont {Watson}}, \bibinfo {author} {\bibfnamefont {Oliver~J.}\
			\bibnamefont {Clark}}, \bibinfo {author} {\bibfnamefont {Federico}\
			\bibnamefont {Mazzola}}, \bibinfo {author} {\bibfnamefont {Igor}\
			\bibnamefont {Markovi\ifmmode~\acute{c}\else \'{c}\fi{}}}, \bibinfo {author}
		{\bibfnamefont {Veronika}\ \bibnamefont {Sunko}}, \bibinfo {author}
		{\bibfnamefont {Timur~K.}\ \bibnamefont {Kim}}, \bibinfo {author}
		{\bibfnamefont {Kai}\ \bibnamefont {Rossnagel}}, \ and\ \bibinfo {author}
		{\bibfnamefont {Philip D.~C.}\ \bibnamefont {King}},\ }\bibfield  {title}
	{\enquote {\bibinfo {title} {{Orbital- and ${k}_{z}$-Selective Hybridization
					of Se $4p$ and Ti $3d$ States in the Charge Density Wave Phase of
					${\mathrm{TiSe}}_{2}$}},}\ }\href {\doibase 10.1103/PhysRevLett.122.076404}
	{\bibfield  {journal} {\bibinfo  {journal} {Phys. Rev. Lett.}\ }\textbf
		{\bibinfo {volume} {122}},\ \bibinfo {pages} {076404} (\bibinfo {year}
		{2019})}\BibitemShut {NoStop}%
	\bibitem [{\citenamefont {Allen}\ and\ \citenamefont
		{Chetty}(1994)}]{Allen1994PRB}%
	\BibitemOpen
	\bibfield  {author} {\bibinfo {author} {\bibfnamefont {Philip~B.}\
			\bibnamefont {Allen}}\ and\ \bibinfo {author} {\bibfnamefont
			{N.}~\bibnamefont {Chetty}},\ }\bibfield  {title} {\enquote {\bibinfo {title}
			{{${\mathrm{TiTe}}_{2}$: Inconsistency between transport properties and
					photoemission results}},}\ }\href {\doibase 10.1103/PhysRevB.50.14855}
	{\bibfield  {journal} {\bibinfo  {journal} {Phys. Rev. B}\ }\textbf {\bibinfo
			{volume} {50}},\ \bibinfo {pages} {14855--14859} (\bibinfo {year}
		{1994})}\BibitemShut {NoStop}%
	\bibitem [{Note1()}]{Note1}%
	\BibitemOpen
	\bibinfo {note} {We compare our resistivity simulations with the sample grown
		at 575 C in Di Salvo 1976, and our Hall effect simulations with the data in
		the same reference on a sample grown at 600 C.}\BibitemShut {Stop}%
	\bibitem [{\citenamefont {Hildebrand}\ \emph {et~al.}(2014)\citenamefont
		{Hildebrand}, \citenamefont {Didiot}, \citenamefont {Novello}, \citenamefont
		{Monney}, \citenamefont {Scarfato}, \citenamefont {Ubaldini}, \citenamefont
		{Berger}, \citenamefont {Bowler}, \citenamefont {Renner},\ and\ \citenamefont
		{Aebi}}]{Hildebrand2014}%
	\BibitemOpen
	\bibfield  {author} {\bibinfo {author} {\bibfnamefont {B.}~\bibnamefont
			{Hildebrand}}, \bibinfo {author} {\bibfnamefont {C.}~\bibnamefont {Didiot}},
		\bibinfo {author} {\bibfnamefont {A.~M.}\ \bibnamefont {Novello}}, \bibinfo
		{author} {\bibfnamefont {G.}~\bibnamefont {Monney}}, \bibinfo {author}
		{\bibfnamefont {A.}~\bibnamefont {Scarfato}}, \bibinfo {author}
		{\bibfnamefont {A.}~\bibnamefont {Ubaldini}}, \bibinfo {author}
		{\bibfnamefont {H.}~\bibnamefont {Berger}}, \bibinfo {author} {\bibfnamefont
			{D.~R.}\ \bibnamefont {Bowler}}, \bibinfo {author} {\bibfnamefont
			{C.}~\bibnamefont {Renner}}, \ and\ \bibinfo {author} {\bibfnamefont
			{P.}~\bibnamefont {Aebi}},\ }\bibfield  {title} {\enquote {\bibinfo {title}
			{{Doping Nature of Native Defects in
					$1T\text{\ensuremath{-}}{\mathrm{TiSe}}_{2}$}},}\ }\href {\doibase
		10.1103/PhysRevLett.112.197001} {\bibfield  {journal} {\bibinfo  {journal}
			{Phys. Rev. Lett.}\ }\textbf {\bibinfo {volume} {112}},\ \bibinfo {pages}
		{197001} (\bibinfo {year} {2014})}\BibitemShut {NoStop}%
	\bibitem [{\citenamefont {Huang}\ \emph {et~al.}(2017)\citenamefont {Huang},
		\citenamefont {Shu}, \citenamefont {Pai}, \citenamefont {Liu},\ and\
		\citenamefont {Chou}}]{Huang2017}%
	\BibitemOpen
	\bibfield  {author} {\bibinfo {author} {\bibfnamefont {S.~H.}\ \bibnamefont
			{Huang}}, \bibinfo {author} {\bibfnamefont {G.~J.}\ \bibnamefont {Shu}},
		\bibinfo {author} {\bibfnamefont {Woei~Wu}\ \bibnamefont {Pai}}, \bibinfo
		{author} {\bibfnamefont {H.~L.}\ \bibnamefont {Liu}}, \ and\ \bibinfo
		{author} {\bibfnamefont {F.~C.}\ \bibnamefont {Chou}},\ }\bibfield  {title}
	{\enquote {\bibinfo {title} {{Tunable Se vacancy defects and the
					unconventional charge density wave in
					$1T\ensuremath{-}{\mathrm{TiSe}}_{2\ensuremath{-}\ensuremath{\delta}}$}},}\
	}\href {\doibase 10.1103/PhysRevB.95.045310} {\bibfield  {journal} {\bibinfo
			{journal} {Phys. Rev. B}\ }\textbf {\bibinfo {volume} {95}},\ \bibinfo
		{pages} {045310} (\bibinfo {year} {2017})}\BibitemShut {NoStop}%
	\bibitem [{\citenamefont {Morosan}\ \emph {et~al.}(2006)\citenamefont
		{Morosan}, \citenamefont {Zandbergen}, \citenamefont {Dennis}, \citenamefont
		{Bos}, \citenamefont {Onose}, \citenamefont {Klimczuk}, \citenamefont
		{Ramirez}, \citenamefont {Ong},\ and\ \citenamefont {Cava}}]{Morosan2006}%
	\BibitemOpen
	\bibfield  {author} {\bibinfo {author} {\bibfnamefont {E.}~\bibnamefont
			{Morosan}}, \bibinfo {author} {\bibfnamefont {H.~W.}\ \bibnamefont
			{Zandbergen}}, \bibinfo {author} {\bibfnamefont {B.~S.}\ \bibnamefont
			{Dennis}}, \bibinfo {author} {\bibfnamefont {J.~W.~G.}\ \bibnamefont {Bos}},
		\bibinfo {author} {\bibfnamefont {Y.}~\bibnamefont {Onose}}, \bibinfo
		{author} {\bibfnamefont {T.}~\bibnamefont {Klimczuk}}, \bibinfo {author}
		{\bibfnamefont {A.~P.}\ \bibnamefont {Ramirez}}, \bibinfo {author}
		{\bibfnamefont {N.~P.}\ \bibnamefont {Ong}}, \ and\ \bibinfo {author}
		{\bibfnamefont {R.~J.}\ \bibnamefont {Cava}},\ }\bibfield  {title} {\enquote
		{\bibinfo {title} {{Superconductivity in Cu$_x$TiSe$_2$}},}\ }\href
	{http://dx.doi.org/10.1038/nphys360} {\bibfield  {journal} {\bibinfo
			{journal} {Nature Physics}\ }\textbf {\bibinfo {volume} {2}},\ \bibinfo
		{pages} {544} (\bibinfo {year} {2006})}\BibitemShut {NoStop}%
	\bibitem [{\citenamefont {Luo}\ \emph {et~al.}(2016)\citenamefont {Luo},
		\citenamefont {Xie}, \citenamefont {Tao}, \citenamefont {Pletikosic},
		\citenamefont {Valla}, \citenamefont {Sahasrabudhe}, \citenamefont
		{Osterhoudt}, \citenamefont {Sutton}, \citenamefont {Burch}, \citenamefont
		{Seibel}, \citenamefont {Krizan}, \citenamefont {Zhu},\ and\ \citenamefont
		{Cava}}]{Luo2016}%
	\BibitemOpen
	\bibfield  {author} {\bibinfo {author} {\bibfnamefont {Huixia}\ \bibnamefont
			{Luo}}, \bibinfo {author} {\bibfnamefont {Weiwei}\ \bibnamefont {Xie}},
		\bibinfo {author} {\bibfnamefont {Jing}\ \bibnamefont {Tao}}, \bibinfo
		{author} {\bibfnamefont {Ivo}\ \bibnamefont {Pletikosic}}, \bibinfo {author}
		{\bibfnamefont {Tonica}\ \bibnamefont {Valla}}, \bibinfo {author}
		{\bibfnamefont {Girija~S.}\ \bibnamefont {Sahasrabudhe}}, \bibinfo {author}
		{\bibfnamefont {Gavin}\ \bibnamefont {Osterhoudt}}, \bibinfo {author}
		{\bibfnamefont {Erin}\ \bibnamefont {Sutton}}, \bibinfo {author}
		{\bibfnamefont {Kenneth~S.}\ \bibnamefont {Burch}}, \bibinfo {author}
		{\bibfnamefont {Elizabeth~M.}\ \bibnamefont {Seibel}}, \bibinfo {author}
		{\bibfnamefont {Jason~W.}\ \bibnamefont {Krizan}}, \bibinfo {author}
		{\bibfnamefont {Yimei}\ \bibnamefont {Zhu}}, \ and\ \bibinfo {author}
		{\bibfnamefont {Robert~J.}\ \bibnamefont {Cava}},\ }\bibfield  {title}
	{\enquote {\bibinfo {title} {{Differences in Chemical Doping Matter:
					Superconductivity in Ti$_{1-x}$Ta$_x$Se$_2$ but Not in
					Ti$_{1-x}$Nb$_x$Se$_2$}},}\ }\href {\doibase 10.1021/acs.chemmater.6b00288}
	{\bibfield  {journal} {\bibinfo  {journal} {Chemistry of Materials}\ }\textbf
		{\bibinfo {volume} {28}},\ \bibinfo {pages} {1927--1935} (\bibinfo {year}
		{2016})}\BibitemShut {NoStop}%
	\bibitem [{\citenamefont {Li}\ \emph {et~al.}(2015)\citenamefont {Li},
		\citenamefont {O'Farrell}, \citenamefont {Loh}, \citenamefont {Eda},
		\citenamefont {{\"O}zyilmaz},\ and\ \citenamefont {Castro~Neto}}]{Li2016}%
	\BibitemOpen
	\bibfield  {author} {\bibinfo {author} {\bibfnamefont {L.~J.}\ \bibnamefont
			{Li}}, \bibinfo {author} {\bibfnamefont {E.~C.~T.}\ \bibnamefont
			{O'Farrell}}, \bibinfo {author} {\bibfnamefont {K.~P.}\ \bibnamefont {Loh}},
		\bibinfo {author} {\bibfnamefont {G.}~\bibnamefont {Eda}}, \bibinfo {author}
		{\bibfnamefont {B.}~\bibnamefont {{\"O}zyilmaz}}, \ and\ \bibinfo {author}
		{\bibfnamefont {A.~H.}\ \bibnamefont {Castro~Neto}},\ }\bibfield  {title}
	{\enquote {\bibinfo {title} {{Controlling many-body states by the
					electric-field effect in a two-dimensional material}},}\ }\href
	{http://dx.doi.org/10.1038/nature16175} {\bibfield  {journal} {\bibinfo
			{journal} {Nature}\ }\textbf {\bibinfo {volume} {529}},\ \bibinfo {pages}
		{185} (\bibinfo {year} {2015})}\BibitemShut {NoStop}%
	\bibitem [{Note2()}]{Note2}%
	\BibitemOpen
	\bibinfo {note} {Note that in Fig.~\ref {fig:fig3}(a) the impurity scattering
		rate is held constant for all carrier densities, although experimentally it
		is likely to scale inversely with $N_e$.}\BibitemShut {Stop}%
	\bibitem [{\citenamefont {Velebit}\ \emph {et~al.}(2016)\citenamefont
		{Velebit}, \citenamefont {Pop\ifmmode \check{c}\else
			\v{c}\fi{}evi\ifmmode~\acute{c}\else \'{c}\fi{}}, \citenamefont
		{Batisti\ifmmode~\acute{c}\else \'{c}\fi{}}, \citenamefont {Eichler},
		\citenamefont {Berger}, \citenamefont {Forr\'o}, \citenamefont {Dressel},
		\citenamefont {Bari\ifmmode \check{s}\else \v{s}\fi{}i\ifmmode~\acute{c}\else
			\'{c}\fi{}},\ and\ \citenamefont {Tuti\ifmmode~\check{s}\else
			\v{s}\fi{}}}]{Velebit2016PRB}%
	\BibitemOpen
	\bibfield  {author} {\bibinfo {author} {\bibfnamefont {K.}~\bibnamefont
			{Velebit}}, \bibinfo {author} {\bibfnamefont {P.}~\bibnamefont {Pop\ifmmode
				\check{c}\else \v{c}\fi{}evi\ifmmode~\acute{c}\else \'{c}\fi{}}}, \bibinfo
		{author} {\bibfnamefont {I.}~\bibnamefont {Batisti\ifmmode~\acute{c}\else
				\'{c}\fi{}}}, \bibinfo {author} {\bibfnamefont {M.}~\bibnamefont {Eichler}},
		\bibinfo {author} {\bibfnamefont {H.}~\bibnamefont {Berger}}, \bibinfo
		{author} {\bibfnamefont {L.}~\bibnamefont {Forr\'o}}, \bibinfo {author}
		{\bibfnamefont {M.}~\bibnamefont {Dressel}}, \bibinfo {author} {\bibfnamefont
			{N.}~\bibnamefont {Bari\ifmmode \check{s}\else
				\v{s}\fi{}i\ifmmode~\acute{c}\else \'{c}\fi{}}}, \ and\ \bibinfo {author}
		{\bibfnamefont {E.}~\bibnamefont {Tuti\ifmmode~\check{s}\else \v{s}\fi{}}},\
	}\bibfield  {title} {\enquote {\bibinfo {title} {{Scattering-dominated
					high-temperature phase of
					$1T\text{\ensuremath{-}}\mathrm{TiS}{\mathrm{e}}_{2}$: An optical
					conductivity study}},}\ }\href {\doibase 10.1103/PhysRevB.94.075105}
	{\bibfield  {journal} {\bibinfo  {journal} {Phys. Rev. B}\ }\textbf {\bibinfo
			{volume} {94}},\ \bibinfo {pages} {075105} (\bibinfo {year}
		{2016})}\BibitemShut {NoStop}%
	\bibitem [{\citenamefont {Chen}\ \emph {et~al.}(2015)\citenamefont {Chen},
		\citenamefont {Chan}, \citenamefont {Fang}, \citenamefont {Zhang},
		\citenamefont {Chou}, \citenamefont {Mo}, \citenamefont {Hussain},
		\citenamefont {Fedorov},\ and\ \citenamefont {Chiang}}]{Chen2015}%
	\BibitemOpen
	\bibfield  {author} {\bibinfo {author} {\bibfnamefont {P.}~\bibnamefont
			{Chen}}, \bibinfo {author} {\bibfnamefont {Y.-H.}\ \bibnamefont {Chan}},
		\bibinfo {author} {\bibfnamefont {X.-Y.}\ \bibnamefont {Fang}}, \bibinfo
		{author} {\bibfnamefont {Y.}~\bibnamefont {Zhang}}, \bibinfo {author}
		{\bibfnamefont {M.~Y.}\ \bibnamefont {Chou}}, \bibinfo {author}
		{\bibfnamefont {S.-K.}\ \bibnamefont {Mo}}, \bibinfo {author} {\bibfnamefont
			{Z.}~\bibnamefont {Hussain}}, \bibinfo {author} {\bibfnamefont {A.-V.}\
			\bibnamefont {Fedorov}}, \ and\ \bibinfo {author} {\bibfnamefont {T.-C.}\
			\bibnamefont {Chiang}},\ }\bibfield  {title} {\enquote {\bibinfo {title}
			{{Charge density wave transition in single-layer titanium diselenide}},}\
	}\href {http://dx.doi.org/10.1038/ncomms9943} {\bibfield  {journal} {\bibinfo
			{journal} {Nature Communications}\ }\textbf {\bibinfo {volume} {6}},\
		\bibinfo {pages} {8943} (\bibinfo {year} {2015})}\BibitemShut {NoStop}%
	\bibitem [{\citenamefont {Kusmartseva}\ \emph {et~al.}(2009)\citenamefont
		{Kusmartseva}, \citenamefont {Sipos}, \citenamefont {Berger}, \citenamefont
		{Forr\'o},\ and\ \citenamefont {Tuti\ifmmode~\check{s}\else
			\v{s}\fi{}}}]{Kusmartseva2009}%
	\BibitemOpen
	\bibfield  {author} {\bibinfo {author} {\bibfnamefont {A.~F.}\ \bibnamefont
			{Kusmartseva}}, \bibinfo {author} {\bibfnamefont {B.}~\bibnamefont {Sipos}},
		\bibinfo {author} {\bibfnamefont {H.}~\bibnamefont {Berger}}, \bibinfo
		{author} {\bibfnamefont {L.}~\bibnamefont {Forr\'o}}, \ and\ \bibinfo
		{author} {\bibfnamefont {E.}~\bibnamefont {Tuti\ifmmode~\check{s}\else
				\v{s}\fi{}}},\ }\bibfield  {title} {\enquote {\bibinfo {title} {{Pressure
					Induced Superconductivity in Pristine
					$1T\mathrm{\text{\ensuremath{-}}}{\mathrm{TiSe}}_{2}$}},}\ }\href {\doibase
		10.1103/PhysRevLett.103.236401} {\bibfield  {journal} {\bibinfo  {journal}
			{Phys. Rev. Lett.}\ }\textbf {\bibinfo {volume} {103}},\ \bibinfo {pages}
		{236401} (\bibinfo {year} {2009})}\BibitemShut {NoStop}%
	\bibitem [{\citenamefont {Morosan}\ \emph {et~al.}(2010)\citenamefont
		{Morosan}, \citenamefont {Wagner}, \citenamefont {Zhao}, \citenamefont {Hor},
		\citenamefont {Williams}, \citenamefont {Tao}, \citenamefont {Zhu},\ and\
		\citenamefont {Cava}}]{Morosan2010PRB}%
	\BibitemOpen
	\bibfield  {author} {\bibinfo {author} {\bibfnamefont {E.}~\bibnamefont
			{Morosan}}, \bibinfo {author} {\bibfnamefont {K.~E.}\ \bibnamefont {Wagner}},
		\bibinfo {author} {\bibfnamefont {Liang~L.}\ \bibnamefont {Zhao}}, \bibinfo
		{author} {\bibfnamefont {Y.}~\bibnamefont {Hor}}, \bibinfo {author}
		{\bibfnamefont {A.~J.}\ \bibnamefont {Williams}}, \bibinfo {author}
		{\bibfnamefont {J.}~\bibnamefont {Tao}}, \bibinfo {author} {\bibfnamefont
			{Y.}~\bibnamefont {Zhu}}, \ and\ \bibinfo {author} {\bibfnamefont {R.~J.}\
			\bibnamefont {Cava}},\ }\bibfield  {title} {\enquote {\bibinfo {title}
			{{Multiple electronic transitions and superconductivity in
					${\text{Pd}}_{x}{\text{TiSe}}_{2}$}},}\ }\href {\doibase
		10.1103/PhysRevB.81.094524} {\bibfield  {journal} {\bibinfo  {journal} {Phys.
				Rev. B}\ }\textbf {\bibinfo {volume} {81}},\ \bibinfo {pages} {094524}
		(\bibinfo {year} {2010})}\BibitemShut {NoStop}%
	\bibitem [{\citenamefont {{Wegner}}\ \emph {et~al.}(2018)\citenamefont
		{{Wegner}}, \citenamefont {{Zhao}}, \citenamefont {{Li}}, \citenamefont
		{{Yang}}, \citenamefont {{Anikin}}, \citenamefont {{Karapetrov}},
		\citenamefont {{Louca}},\ and\ \citenamefont
		{{Chatterjee}}}]{Wegner2018_arxiv}%
	\BibitemOpen
	\bibfield  {author} {\bibinfo {author} {\bibfnamefont {A.}~\bibnamefont
			{{Wegner}}}, \bibinfo {author} {\bibfnamefont {J.}~\bibnamefont {{Zhao}}},
		\bibinfo {author} {\bibfnamefont {J.}~\bibnamefont {{Li}}}, \bibinfo {author}
		{\bibfnamefont {J.}~\bibnamefont {{Yang}}}, \bibinfo {author} {\bibfnamefont
			{A.~A.}\ \bibnamefont {{Anikin}}}, \bibinfo {author} {\bibfnamefont
			{G.}~\bibnamefont {{Karapetrov}}}, \bibinfo {author} {\bibfnamefont
			{D.}~\bibnamefont {{Louca}}}, \ and\ \bibinfo {author} {\bibfnamefont
			{U.}~\bibnamefont {{Chatterjee}}},\ }\bibfield  {title} {\enquote {\bibinfo
			{title} {{Evidence for breathing-type pseudo Jahn-Teller distortions in the
					charge density wave phase of 1$T$-TiSe$_2$}},}\ }\href
	{https://arxiv.org/abs/1807.05664} {\bibfield  {journal} {\bibinfo  {journal}
			{arXiv: 1807.05664}\ } (\bibinfo {year} {2018})}\BibitemShut {NoStop}%
	\bibitem [{\citenamefont {Castellan}\ \emph {et~al.}(2013)\citenamefont
		{Castellan}, \citenamefont {Rosenkranz}, \citenamefont {Osborn},
		\citenamefont {Li}, \citenamefont {Gray}, \citenamefont {Luo}, \citenamefont
		{Welp}, \citenamefont {Karapetrov}, \citenamefont {Ruff},\ and\ \citenamefont
		{van Wezel}}]{Castellan2013}%
	\BibitemOpen
	\bibfield  {author} {\bibinfo {author} {\bibfnamefont {John-Paul}\
			\bibnamefont {Castellan}}, \bibinfo {author} {\bibfnamefont {Stephan}\
			\bibnamefont {Rosenkranz}}, \bibinfo {author} {\bibfnamefont {Ray}\
			\bibnamefont {Osborn}}, \bibinfo {author} {\bibfnamefont {Qing'an}\
			\bibnamefont {Li}}, \bibinfo {author} {\bibfnamefont {K.~E.}\ \bibnamefont
			{Gray}}, \bibinfo {author} {\bibfnamefont {X.}~\bibnamefont {Luo}}, \bibinfo
		{author} {\bibfnamefont {U.}~\bibnamefont {Welp}}, \bibinfo {author}
		{\bibfnamefont {Goran}\ \bibnamefont {Karapetrov}}, \bibinfo {author}
		{\bibfnamefont {J.~P.~C.}\ \bibnamefont {Ruff}}, \ and\ \bibinfo {author}
		{\bibfnamefont {Jasper}\ \bibnamefont {van Wezel}},\ }\bibfield  {title}
	{\enquote {\bibinfo {title} {{Chiral Phase Transition in Charge Ordered
					$1T\mathrm{\text{\ensuremath{-}}}{\mathrm{TiSe}}_{2}$}},}\ }\href {\doibase
		10.1103/PhysRevLett.110.196404} {\bibfield  {journal} {\bibinfo  {journal}
			{Phys. Rev. Lett.}\ }\textbf {\bibinfo {volume} {110}},\ \bibinfo {pages}
		{196404} (\bibinfo {year} {2013})}\BibitemShut {NoStop}%
	\bibitem [{\citenamefont {Li}\ \emph {et~al.}(2016)\citenamefont {Li},
		\citenamefont {Zhao}, \citenamefont {Liu}, \citenamefont {Ren}, \citenamefont
		{Eda},\ and\ \citenamefont {Loh}}]{Li2016APL}%
	\BibitemOpen
	\bibfield  {author} {\bibinfo {author} {\bibfnamefont {L.~J.}\ \bibnamefont
			{Li}}, \bibinfo {author} {\bibfnamefont {W.~J.}\ \bibnamefont {Zhao}},
		\bibinfo {author} {\bibfnamefont {B.}~\bibnamefont {Liu}}, \bibinfo {author}
		{\bibfnamefont {T.~H.}\ \bibnamefont {Ren}}, \bibinfo {author} {\bibfnamefont
			{G.}~\bibnamefont {Eda}}, \ and\ \bibinfo {author} {\bibfnamefont {K.~P.}\
			\bibnamefont {Loh}},\ }\bibfield  {title} {\enquote {\bibinfo {title}
			{{Enhancing charge-density-wave order in 1T-TiSe$_2$ nanosheet by
					encapsulation with hexagonal boron nitride}},}\ }\href {\doibase
		10.1063/1.4963885} {\bibfield  {journal} {\bibinfo  {journal} {Applied
				Physics Letters}\ }\textbf {\bibinfo {volume} {109}},\ \bibinfo {pages}
		{141902} (\bibinfo {year} {2016})},\ \Eprint
	{http://arxiv.org/abs/https://doi.org/10.1063/1.4963885}
	{https://doi.org/10.1063/1.4963885} \BibitemShut {NoStop}%
	\bibitem [{\citenamefont {Kogar}\ \emph {et~al.}(2017)\citenamefont {Kogar},
		\citenamefont {de~la Pena}, \citenamefont {Lee}, \citenamefont {Fang},
		\citenamefont {Sun}, \citenamefont {Lioi}, \citenamefont {Karapetrov},
		\citenamefont {Finkelstein}, \citenamefont {Ruff}, \citenamefont
		{Abbamonte},\ and\ \citenamefont {Rosenkranz}}]{Kogar2017PRL}%
	\BibitemOpen
	\bibfield  {author} {\bibinfo {author} {\bibfnamefont {A.}~\bibnamefont
			{Kogar}}, \bibinfo {author} {\bibfnamefont {G.~A.}\ \bibnamefont {de~la
				Pena}}, \bibinfo {author} {\bibfnamefont {Sangjun}\ \bibnamefont {Lee}},
		\bibinfo {author} {\bibfnamefont {Y.}~\bibnamefont {Fang}}, \bibinfo {author}
		{\bibfnamefont {S.~X.-L.}\ \bibnamefont {Sun}}, \bibinfo {author}
		{\bibfnamefont {D.~B.}\ \bibnamefont {Lioi}}, \bibinfo {author}
		{\bibfnamefont {G.}~\bibnamefont {Karapetrov}}, \bibinfo {author}
		{\bibfnamefont {K.~D.}\ \bibnamefont {Finkelstein}}, \bibinfo {author}
		{\bibfnamefont {J.~P.~C.}\ \bibnamefont {Ruff}}, \bibinfo {author}
		{\bibfnamefont {P.}~\bibnamefont {Abbamonte}}, \ and\ \bibinfo {author}
		{\bibfnamefont {S.}~\bibnamefont {Rosenkranz}},\ }\bibfield  {title}
	{\enquote {\bibinfo {title} {{Observation of a Charge Density Wave
					Incommensuration Near the Superconducting Dome in
					${\mathrm{Cu}}_{x}{\mathrm{TiSe}}_{2}$}},}\ }\href {\doibase
		10.1103/PhysRevLett.118.027002} {\bibfield  {journal} {\bibinfo  {journal}
			{Phys. Rev. Lett.}\ }\textbf {\bibinfo {volume} {118}},\ \bibinfo {pages}
		{027002} (\bibinfo {year} {2017})}\BibitemShut {NoStop}%
	\bibitem [{\citenamefont {Li}\ \emph {et~al.}(2007)\citenamefont {Li},
		\citenamefont {Hu}, \citenamefont {Qian}, \citenamefont {Hsieh},
		\citenamefont {Hasan}, \citenamefont {Morosan}, \citenamefont {Cava},\ and\
		\citenamefont {Wang}}]{Li2007}%
	\BibitemOpen
	\bibfield  {author} {\bibinfo {author} {\bibfnamefont {G.}~\bibnamefont
			{Li}}, \bibinfo {author} {\bibfnamefont {W.~Z.}\ \bibnamefont {Hu}}, \bibinfo
		{author} {\bibfnamefont {D.}~\bibnamefont {Qian}}, \bibinfo {author}
		{\bibfnamefont {D.}~\bibnamefont {Hsieh}}, \bibinfo {author} {\bibfnamefont
			{M.~Z.}\ \bibnamefont {Hasan}}, \bibinfo {author} {\bibfnamefont
			{E.}~\bibnamefont {Morosan}}, \bibinfo {author} {\bibfnamefont {R.~J.}\
			\bibnamefont {Cava}}, \ and\ \bibinfo {author} {\bibfnamefont {N.~L.}\
			\bibnamefont {Wang}},\ }\bibfield  {title} {\enquote {\bibinfo {title}
			{{Semimetal-to-Semimetal Charge Density Wave Transition in
					$1T\mathrm{\text{\ensuremath{-}}}{\mathrm{TiSe}}_{2}$}},}\ }\href {\doibase
		10.1103/PhysRevLett.99.027404} {\bibfield  {journal} {\bibinfo  {journal}
			{Phys. Rev. Lett.}\ }\textbf {\bibinfo {volume} {99}},\ \bibinfo {pages}
		{027404} (\bibinfo {year} {2007})}\BibitemShut {NoStop}%
	\bibitem [{\citenamefont {May}\ \emph {et~al.}(2011)\citenamefont {May},
		\citenamefont {Brabetz}, \citenamefont {Janowitz},\ and\ \citenamefont
		{Manzke}}]{May2011PRL}%
	\BibitemOpen
	\bibfield  {author} {\bibinfo {author} {\bibfnamefont {Matthias~M.}\
			\bibnamefont {May}}, \bibinfo {author} {\bibfnamefont {Christine}\
			\bibnamefont {Brabetz}}, \bibinfo {author} {\bibfnamefont {Christoph}\
			\bibnamefont {Janowitz}}, \ and\ \bibinfo {author} {\bibfnamefont {Recardo}\
			\bibnamefont {Manzke}},\ }\bibfield  {title} {\enquote {\bibinfo {title}
			{{Charge-Density-Wave Phase of
					$1T\mathrm{\text{\ensuremath{-}}}{\mathrm{TiSe}}_{2}$: The Influence of
					Conduction Band Population}},}\ }\href {\doibase
		10.1103/PhysRevLett.107.176405} {\bibfield  {journal} {\bibinfo  {journal}
			{Phys. Rev. Lett.}\ }\textbf {\bibinfo {volume} {107}},\ \bibinfo {pages}
		{176405} (\bibinfo {year} {2011})}\BibitemShut {NoStop}%
	\bibitem [{\citenamefont {Hellgren}\ \emph {et~al.}(2017)\citenamefont
		{Hellgren}, \citenamefont {Baima}, \citenamefont {Bianco}, \citenamefont
		{Calandra}, \citenamefont {Mauri},\ and\ \citenamefont
		{Wirtz}}]{Hellgren2017}%
	\BibitemOpen
	\bibfield  {author} {\bibinfo {author} {\bibfnamefont {Maria}\ \bibnamefont
			{Hellgren}}, \bibinfo {author} {\bibfnamefont {Jacopo}\ \bibnamefont
			{Baima}}, \bibinfo {author} {\bibfnamefont {Raffaello}\ \bibnamefont
			{Bianco}}, \bibinfo {author} {\bibfnamefont {Matteo}\ \bibnamefont
			{Calandra}}, \bibinfo {author} {\bibfnamefont {Francesco}\ \bibnamefont
			{Mauri}}, \ and\ \bibinfo {author} {\bibfnamefont {Ludger}\ \bibnamefont
			{Wirtz}},\ }\bibfield  {title} {\enquote {\bibinfo {title} {{Critical Role of
					the Exchange Interaction for the Electronic Structure and Charge-Density-Wave
					Formation in ${\mathrm{TiSe}}_{2}$}},}\ }\href {\doibase
		10.1103/PhysRevLett.119.176401} {\bibfield  {journal} {\bibinfo  {journal}
			{Phys. Rev. Lett.}\ }\textbf {\bibinfo {volume} {119}},\ \bibinfo {pages}
		{176401} (\bibinfo {year} {2017})}\BibitemShut {NoStop}%
\end{thebibliography}

%

\end{document}